\documentclass[a4paper]{jpconf}
\usepackage{siunitx}
\usepackage{graphicx}
\usepackage{subcaption}

\usepackage{hyperref}
\hypersetup{
    colorlinks=true,
    citecolor=blue,
    hidelinks,
}

\usepackage[sort&compress,numbers]{natbib}

\newcommand{\isot}[2]{\ensuremath{{}^{#2}\mathrm{#1}}}

\newcommand{\castro}{{\sf Castro}}
\newcommand{\amrex}{{\sf AMReX}}

\newcommand{\rprox}{\texttt{rprox}}

\long\def\software{\bgroup \@software {[}}
\def\@software[#1]#2{\vskip 6pt{
\frenchspacing

#2
}\egroup}

\begin{document}
\title{Simulating Lateral H/He Flame Propagation in Type I X-ray Bursts}

\author{Eric T. Johnson$^1$,
        Michael Zingale$^1$}

\address{$^1$Department of Physics and Astronomy, Stony Brook
  University, Stony Brook, NY 11794-3800 USA}

\ead{eric.t.johnson@stonybrook.edu}

\begin{abstract}
X-ray bursts are the thermonuclear runaway of a mixed H/He layer on the surface
of a neutron star.  Observations suggest that the burning begins locally and spreads
across the surface of the star as a flame.  Recent multidimensional 
work has looked in detail at pure He flames spreading across a neutron star.  Here
we report on progress in multidimensional modeling of mixed H/He flames
and discuss the challenges.
\end{abstract}

\section{Introduction}

Type I X-ray bursts (XRBs) occur in accreting binary systems between a neutron star and a low-mass companion.
As a layer of fuel builds up on the surface of the neutron star, gravitational compression heats it up until the conditions at the base lead to a thermonuclear runaway.
A challenging aspect of modeling these systems is capturing how the flame spreads from the initial ignition point across the surface of the star.
This is important for understanding oscillations observed during the initial rise of some bursts, which are believed to arise from the bright hotspot coming in and out of view as the neutron star rotates \cite{chakraborty:2014}.
Unfortunately, one-dimensional simulations cannot capture these dynamics, as they are inherently multi-dimensional.

Most burst sources accrete a mix of \isot{H}{1} and \isot{He}{4} from a main sequence companion, while some ultra-compact systems accrete nearly pure \isot{He}{4} \cite{galloway:2017,intZand:2007}.
Depending on the accretion rate, the hydrogen can burn stably into helium which will then ignite a pure helium burst, or the helium could ignite before all the hydrogen is consumed, leading to a mixed H/He burst.
Previous 2D and 3D simulations by our group and others have been limited to pure helium flames \cite{cavecchi:2013,art-2015-cavecchi-etal,Cavecchi2019,eiden:2020,harpole:2021,goodwin:2021,zingale:2023} as the mixed H/He reaction networks are usually larger and more difficult to integrate, and the timescales are slower due to the weak rate waiting points \cite{STRO_BILD06}.
We recently showed \cite{zingale:2023} that the nucleosynthesis and evolution of a spreading flame in 3D matches 2D simulations very closely, so we will focus on 2D simulations here.

\section{Simulations}
We use the \castro~\cite{castro} compressible hydrodynamics code to perform our simulations, following the simulation methodology developed in \cite{eiden:2020}.
\castro{} uses the \amrex{} framework \cite{amrex_joss} for adaptive mesh refinement and parallelization across CPUs and GPUs.
We use the stellar equation of state based on \cite{timmes_swesty:2000} and conductivities from \cite{Timmes00} for the thermal diffusion operator.
\castro\ can use an arbitrary nuclear reaction network, and we discuss some of the options below.

All simulations use a two-dimensional axisymmetric simulation domain with a physical size of \qtyproduct{2.4576e5 x 3.072e4}{\cm} and a base grid of \numproduct{1024 x 128} zones.
The grid is subdivided using two levels of refinement, refining first by a factor of 4, then by a factor of 2.
This gives a resolution of \qty{30}{\cm} at the finest level.
We note that this is a little coarser than the He simulations we have done previously.
However, since the addition of H to the atmosphere increases the scale height and H burning is less strongly peaked than He burning, we believe we can relax the resolution requirements.

\begin{figure}[t]
\begin{subfigure}{0.49\linewidth}
    \includegraphics{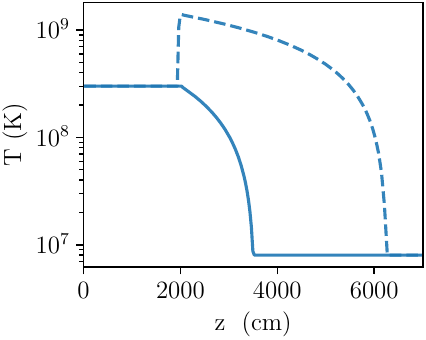}
    \caption{\rprox\ run with 74\% H, 25\% He}
    \label{fig:initial-model-rprox}
\end{subfigure}
\begin{subfigure}{0.49\linewidth}
    \includegraphics{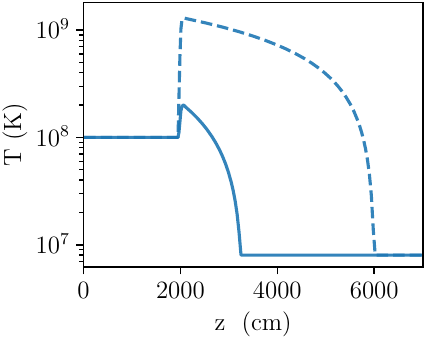}
    \caption{\texttt{aprox19} run with 73\% H, 25\% He}
    \label{fig:initial-model-aprox19}
\end{subfigure}
\caption{Initial vertical temperature profiles for the cool (solid) and hot (dashed) regions.}
\label{fig:initial-models}
\end{figure}

Our initial runs used \rprox, a 10 isotope nuclear reaction network introduced in \cite{wallacewoosley:1981} that models hot CNO burning, triple-$\alpha$, and rp-process breakout.
We vary the proportions of \isot{H}{1} and \isot{He}{4} in the fuel layer, and also include 1\% of \isot{O}{14} and \isot{O}{15} to help start CNO burning.
The oxygen abundance is split between the two isotopes in proportion to their respective $\beta^+$-decay lifetimes.
We assume that the composition is completely mixed by a period of convection preceding the ignition of the flame, as explored in \cite{xrb2}.

The initial model is set up as described in section 4 of \cite{eiden:2020}.
In summary, it consists of a hot region behind the flame at $T=\qty{1.4e9}{\kelvin}$ and a cool region ahead of it at $T=\qty{3e8}{\kelvin}$, with a density of $\qty{2e6}{\gram\per\cm\cubed}$ at the base of the atmosphere ($z=\qty{2000}{\cm}$).
These two regions are constructed individually in 1D to be in hydrostatic equilibrium vertically, then smoothly blended at the interface, giving us an equilibrium starting condition that will lead to the ignition of a burning front.
Figure~\ref{fig:initial-model-rprox} shows the temperature profiles for the two regions for one of the runs. The other profiles with different compositions are similar, but those with more helium are squashed due to the greater mass of the atmosphere above the surface.
The hot region extends to $r_\text{pert}=\qty{6.144e4}{\cm}$ (25\% of the domain), and the transition interface between the two regions has a width of $\delta_\text{blend}=\qty{2.048e3}{\cm}$.
Paired with the axisymmetric geometry, this effectively models a hotspot at the pole of the star spreading outwards towards the equator.
The rotation rate of the neutron star is set to an artificially high \qty{1000}{\hertz}, which helps to confine the flame laterally and allows us to use a smaller simulation domain.
All of our simulations were run on the OLCF Summit machine, with the entire calculation offloaded to the NVIDIA V100 GPUs there.

\begin{figure}
\centering
\includegraphics{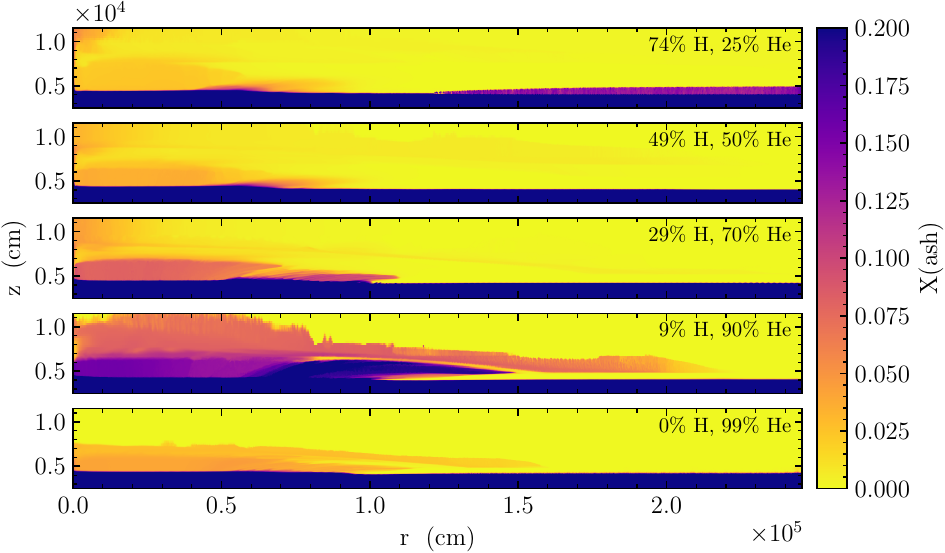}
\caption{Total mass fraction of all species heavier than \isot{O}{16} for \rprox\ runs at $t = \qty{80}{\ms}$.}
\label{fig:rprox-comparison}
\end{figure}

Figure~\ref{fig:rprox-comparison} shows snapshots of the mass fraction of ash (everything heavier than \isot{O}{16}) after \qty{80}{\ms} for each of the initial compositions we used.
We have found that this diagnostic is much easier to compare between the different compositions than the mean molecular weight, $\bar{A}$, used in our previous work.
The runs in the first two panels never produced well-defined flames.
The third panel has more burning than the previous two, but the flame front never moves beyond \qty{e5}{\cm}.
The most interesting is the fourth panel, with 9\% H and 90\% He, which shows a flame front with burning up to \isot{Ni}{56}.
The final panel models a pure He burst using the same network for comparison.

Many of our runs exhibited an unexpected high-temperature region at the base of the atmosphere across the entire simulation domain, starting from the outer radial edge of the domain and spreading toward the pole.
This often ignited and consumed all the fuel before the flame could set up and start propagating.
The magenta layer on the right of the first panel in figure~\ref{fig:rprox-comparison}, for example, is a result of this excess burning.
We attributed this to our selection of the base temperature in the cool region, and used a lower temperature in later runs to try to avoid this.
This is one of the main differences between a pure He burst---a pure He network will not burn much at those temperatures.
However, we found that the simulation behavior was very sensitive to any changes in the initial conditions, and it was difficult to reproduce a steady burning front.
To explore this behavior in more detail, we experimented with 1D flame simulations using different reaction networks, and found that \texttt{aprox19} produced a much more robust propagating flame front than \rprox.
Due to the different composition in that network, we replaced the oxygen isotopes with \isot{C}{12}.

\begin{figure}
\centering
\includegraphics{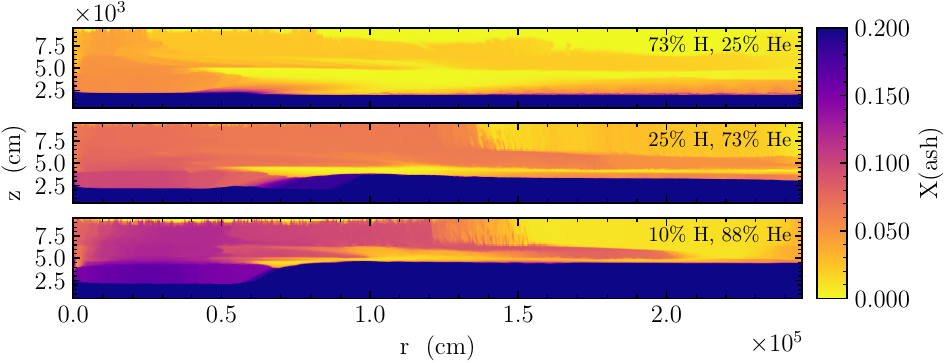}
\caption{Total ash mass fraction at $t = \qty{80}{\ms}$ for \texttt{aprox19} runs.}
\label{fig:aprox19-comparison}
\end{figure}

For our next set of runs with \texttt{aprox19}, we increased the resolution to \qty{20}{\cm} (using a base grid of \numproduct{1536 x 192}) and used a new lower boundary condition with a well-balanced pressure reconstruction, as done in \cite{zingale:2023}.
We reduced the cool region temperature to \qty{2e8}{\kelvin} and the material below the atmosphere (i.e. the star) to \qty{1e8}{\kelvin}, as mentioned above.
The new initial conditions are shown in figure~\ref{fig:initial-model-aprox19}.
Interestingly, increasing the resolution this way actually scaled slightly better than linear in the number of compute nodes, as the new grid size distributed more evenly onto Summit's 6 GPUs per node.

\begin{figure}
\centering
\includegraphics{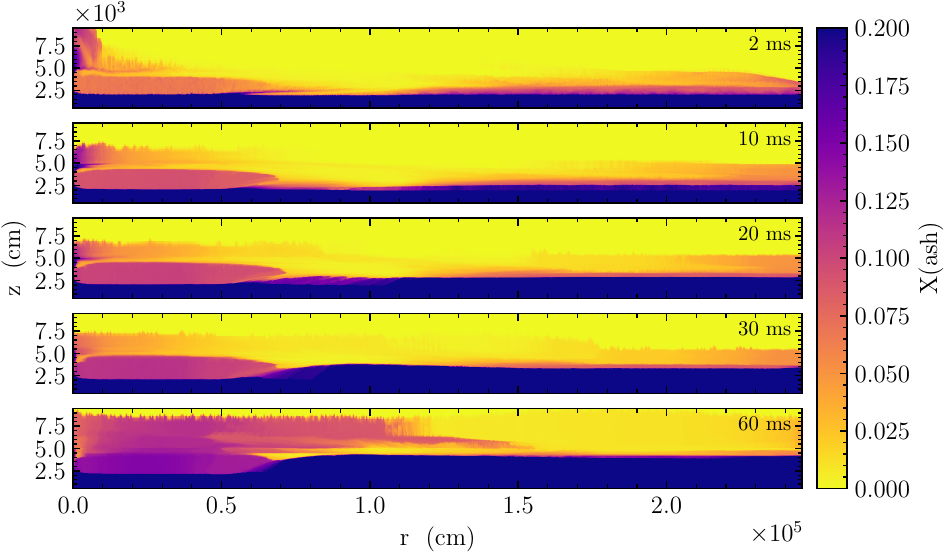}
\caption{Ash mass fraction evolution for the \texttt{aprox19} run with 10\% H and 88\% He.}
\label{fig:aprox19-time-series}
\end{figure}

Figure~\ref{fig:aprox19-comparison} shows the same diagnostics as figure~\ref{fig:rprox-comparison} for the \texttt{aprox19} runs.
We can see that the run with 25\% He again doesn't produce a flame, although it does have more ash present in the atmosphere than the corresponding \rprox\ run.
This is due to vigorous carbon burning in the first few milliseconds of the simulation, which consumes the majority of the initial \isot{C}{12} in the lower atmosphere.
In other two runs, the right side of the atmosphere heats up and starts burning before the flame front.
An example of how this progresses for the run with 10\% H and 88\% He is shown in figure~\ref{fig:aprox19-time-series}.

\begin{figure}
\centering
\includegraphics{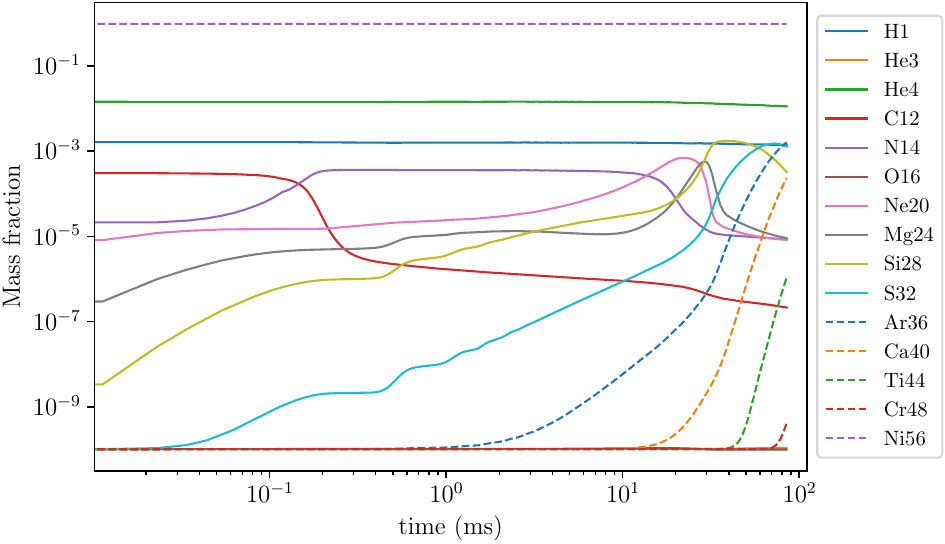}
\caption{Total mass fractions for each species in the \texttt{aprox19} run with 10\% H, excluding those which never increased above \num{e-10}.}
\label{fig:aprox19-mass-fractions}
\end{figure}

Finally, figure~\ref{fig:aprox19-mass-fractions} shows the evolution of each species in the network over time, for the same run.
We can see that most of the initial \isot{C}{12} (solid red line) is consumed in under 1 millisecond.
An interesting detail in this plot is the step-like features at early times, most visible in the profiles for \isot{Si}{28} and \isot{S}{32}.
These are fairly evenly spaced in time with a separation of $\mathop{\sim}\qty{5e-4}{\second}$, which matches the width of the domain divided by the average sound speed near the base of the atmosphere: $L_x / c_s \approx \qty{2.4576e5}{\cm} / \qty{5.2e8}{\cm\per\second} \approx \qty{4.7e-4}{\second}$.
This suggests they may be related to acoustic oscillations travelling across the simulation domain and reflecting off of the boundaries.

\section{Summary}

We ran several two-dimensional simulations of mixed H/He flames in XRBs, varying the proportions of H and He in the atmosphere.
We found that the flames are very sensitive to the composition and initial temperatures, which makes it difficult to set up a flame in the first place.
Using the \rprox\ network, we were only able to produce a stable, propagating flame with an initial composition of 9\% H and 90\% He.
The \texttt{aprox19} network gave much better results in 1D simulations, but we have not yet been able to reproduce a clean H/He flame in 2D.

An unexpected hot region would often develop at the surface of the star, away from the flame, which would start burning and disrupt the flame.
The source of this heating is unclear, and needs to be investigated further.
We are currently testing out different reaction networks with pynucastro \cite{pynucastro,pynucastro2}, which allows us to construct and inspect arbitrary reaction networks, and generate C++ code for use in our \castro{} simulations.
It also allows us to disable specific rates at runtime, which can help us understand which rates are relevant for H/He burning under these conditions.
Finally, we may need to
consider different initial models, where
the pre-runaway convective mixing is confined only to the He layer at high column depth, as explored in \cite{guichandut:2023}.

\ack  \castro\ is freely available at
\url{http://github.com/AMReX-Astro/Castro}.  All of the code and
problem setups used here are available in the git repo.  The work at
Stony Brook was supported by DOE/Office of Nuclear Physics grant
DE-FG02-87ER40317.  This research used resources of the Oak Ridge Leadership Computing
Facility at the Oak Ridge National Laboratory, which is supported by
the Office of Science of the U.S. Department of Energy under Contract
No.\ DE-AC05-00OR22725, awarded through the DOE INCITE program.  We
thank NVIDIA Corporation for the donation of a Titan X and Titan V GPU
through their academic grant program.  This research has made use of
NASA's Astrophysics Data System Bibliographic Services.

{\it Software: }%
\amrex~\citep{amrex_joss},
\castro~\citep{castro,castro_joss},
GCC (\url{https://gcc.gnu.org/}),
GNU Parallel \citep{tange:2018},
helmeos \citep{timmes_swesty:2000},
linux (\url{https://www.kernel.org/}),
matplotlib \citep{Hunter:2007},
NumPy \citep{numpy,numpy2},
python (\url{https://www.python.org/}),
valgrind \citep{valgrind},
VODE \citep{vode},
yt \citep{yt}.

\bibliographystyle{iopart-num}
\bibliography{ws}

\end{document}